\documentclass{aa}
\usepackage{graphicx}
\usepackage{natbib}
\usepackage{amssymb}
\bibpunct{(}{)}{;}{a}{}{,}
\begin{document}
   \title{Properties of detached shells around carbon stars}

   \subtitle{ Evidence of interacting winds}

   \author{F. L. Sch\"oier\inst{1}  \and M. Lindqvist\inst{2}  \and H. Olofsson\inst{1}}

   \offprints{F. L. Sch\"oier \\ \email{fredrik@astro.su.se}}

   \institute{Stockholm Observatory, AlbaNova, SE-106 91 Stockholm, Sweden
   \and Onsala Space Observatory, SE-439 92 Onsala, Sweden}    
             
   \date{Received; accepted}

   \abstract{The nature of the mechanism responsible for producing the spectacular, geometrically thin, spherical shells found around some carbon stars has been an enigma for some time. Based on extensive radiative transfer modelling of both CO line emission and dust continuum radiation for all objects with known detached molecular shells, we present compelling evidence that these shells show clear signs of interaction with a surrounding medium. The derived masses of the shells increase with radial distance from the central star while their velocities decrease. A simple model for interacting winds indicate that the  mass-loss rate producing the faster moving wind has to be almost two orders of magnitudes higher ($\sim$\,10$^{-5}$\,M$_{\odot}$\,yr$^{-1}$) than the slower AGB wind (a few 10$^{-7}$\,M$_{\odot}$\,yr$^{-1}$) preceding this violent event. At the same time, the present-day mass-loss rates are very low indicating that the epoch of high  mass-loss rate was relatively short, on the order of a few hundred years. This, together with the number of sources exhibiting this phenomenon, suggests a connection with He-shell flashes (thermal pulses). 
   We report the detection of a detached molecular shell around the carbon star \object{DR~Ser}, as revealed from new single-dish CO (sub-)millimetre line observations. The properties of the shell are similar to those characterising the young shell around \object{U~Cam}. 
      
   \keywords{Stars: AGB and post-AGB -- Stars: carbon -- Stars: late-type -- Stars: mass-loss}
   }
   \maketitle
%

\section{Introduction}
The late evolutionary stages of low- to intermediate-mass stars, as
they ascend the asymptotic giant branch (AGB), are characterized by the ejection of
gas and dust via a slow (5\,$-$\,30\,km\,s$^{-1}$) stellar wind. This mass loss (on the
order of $10^{-7}$\,--\,$10^{-5}$\,M$_{\odot}$~yr$^{-1}$) is crucial
to the evolution of the star and a key process
for enriching and replenishing the interstellar medium \citep[e.g.,][]{Schroeder99, Schroeder01}. Yet, the
mechanisms by which it occurs are not understood in detail.  In
particular, there is evidence that strong variations in the mass-loss
rate, perhaps related to flashes of helium shell burning \citep{Olofsson90}, 
can lead to the formation of circumstellar detached shells. The existence of
such shells can be inferred from their excess emission at
far-infrared wavelengths (due to a lack of hot dust close to the star,
\citealt{Willems88,Zijlstra92}), and are confirmed by a ring-like morphology in maps
of dust or CO line emission \citep[e.g.,][]{Waters94,Olofsson96,Lindqvist99,Olofsson00}.

Detached CO shells have been observed around about a half-dozen AGB stars
\citep[see review by][]{Wallerstein98}, all of them carbon stars. However, 
only \object{U~Cam}
\citep{Lindqvist96,Lindqvist99}, \object{TT~Cyg}
\citep{Olofsson00}, and \object{S~Sct} (Olofsson et al., in prep.) have been mapped in detail with millimetre-wave interferometers. 
These three cases reveal remarkably symmetric shells
expanding away from the stars. However, there are also clear indications of a highly clumped medium.  High-resolution molecular
line observations offer the possibility not only to confirm
detachment, but also to study variations in mass loss on short ($<$100
yr) timescales, related to the shell thickness, and to look for
departures from spherical symmetry. Single-dish CO maps provide evidence for
detached CO shells also in the cases of \object{R~Scl}, \object{U~Ant}, and
\object{V644~Sco} \citep{Olofsson96}.

\citet{Delgado01,Delgado03b} have imaged the circumstellar media of \object{R~Scl}
and \object{U~Ant} in circumstellar scattered stellar light. They found evidence 
for both dust- and gas-scattered light in detached shells of sizes comparable
to those found in the CO line emission. These data provide
high-angular-resolution information, but the separation of dust- and gas-scattered
light is problematic.

Based on colour-colour diagrams for a large sample of OH/IR stars  \citet{Lewis04} have recently suggested that the vast majority of them could have detached shells, possibly linked to a brief period of intense mass loss for oxygen-rich AGB stars \citep{Lewis02}. However, no detached CO shell has been found for any oxygen-rich object yet.

Evidence of detached dust shells have been found in the cases of \object{U~Ant}
\citep{Izumiura97}, \object{U~Hya} \citep{Waters94}, \object{Y~CVn} \citep{Izumiura96},
and \object{R~Hya} \citep{Hashimoto98}. All, but the last are carbon stars.
Except for \object{U~Ant}, there has been no detections of detached CO shells for
these stars. 

\citet{Olofsson90} suggested that the detached shells were an effect of
a mass-loss-rate modulation caused by a He-shell flash (i.e., a thermal pulse).
\citet{Vassiliadis93} studied this in some more detail, and followed the total mass-loss 
history of an AGB star. More recently, \citet{Schroeder98}, \citet{Schroeder99} and 
\citet{Wachter02} combined stellar
evolutionary models and mass-loss prescriptions to study the effects on the 
evolution of an AGB star. They argued that during the low-mass-loss-rate evolution only
carbon stars reach a critical (Eddington-like) luminosity that drives an intense
mass ejection during a He-shell flash, hence explaining the absence (or at least rare occurrence) of detached shells towards M-type AGB
stars. At higher mass-loss rates the detached shells become, in comparison, less
conspicuous. \citet{Steffen00} used hydrodynamical simulations to follow the
circumstellar evolution due to changes in the stellar mass-loss rate. They verified that
a brief period of very high mass-loss rate translates into an expanding, geometrically
thin shell around the star. They also concluded that an interacting-wind scenario
is an equally viable explanation for the detached shells. 

Less dramatic modulations of the mass-loss rate on shorter time scales have been observed towards the archetypical high-mass-loss-rate carbon star \object{IRC+10216} \citep{Mauron99,Mauron00,Fong03}.
Thin arcs, produced by mass-loss-rate modulations while the central star was ascending the AGB, are also seen around PPNe \citep{Hrivnak01,Su03} and PNe \citep{Corradi04}.

In this paper we present new single-dish millimetre line observations of \object{R~Scl}, \object{U~Cam}, \object{V644~Sco},  and \object{DR~Ser}, Sect.~2. Of
particular importance is 
the detection of a new carbon star with a young detached molecular shell, \object{DR~Ser}, increasing the number of such stars to seven. 
In fact, there is not much hope of increasing this number further since most of
the reasonably nearby AGB stars with mass loss have already been searched for
circumstellar CO radio line emission. 
Section~3 describes
the analysis of the molecular line and dust emission. Properties of the detached shells (Sect.~4), and the present-day winds (Sect.~5) of seven carbon stars with
detached molecular shells, Table~\ref{radio}, are estimated. The results are discussed in Sect.~6, where different scenarios for producing the shells are reviewed. The usefulness of future high spatial resolution, as well as high-$J$ CO observations, in further constraining the shell characteristics are also discussed. Our conclusions are presented in Sect.~7.


%
\begin{table}
\caption[]{Carbon stars with detached molecular shells.}
\label{radio}
$$
\begin{array}{p{0.3\linewidth}lclccl}
\hline
\noalign{\smallskip}
\multicolumn{1}{l}{{\mathrm{Source}}} &
\multicolumn{1}{c}{{\mathrm{RA}}(2000.0)} & &
\multicolumn{1}{c}{{\mathrm{DEC}}(2000.0)} &&
\multicolumn{1}{c}{{\mathrm{Ref.}}} 
  \\
&
\multicolumn{1}{c}{[\mathrm{h:m:s}]} & &
\multicolumn{1}{c}{[{\degr:\arcmin:\arcsec}]} &&
\\
\noalign{\smallskip}
\hline
\noalign{\smallskip}
R Scl         & \mbox{01:26:58.09} & & -\mbox{32:32:35.5} && 1\\
U Cam      & \mbox{03:41:48.17} & & \phantom{-}\mbox{62:38:54.5} && 2\\
U Ant         & \mbox{10:35:12.85} & & -\mbox{39:33:45.3} & & 3 \\
V644 Sco  &\mbox{17:26:18.7}   & & -\mbox{40:01:52} & & 4\\
DR Ser      & \mbox{18:47:21.02} & & \phantom{-}\mbox{05:27:18.6} & & 3 \\
S Sct         & \mbox{18:50:20.04} & & -\mbox{07:54:27.4} & & 3\\
TT Cyg     & \mbox{19:40:57.00} & & \phantom{-}\mbox{32:37:05.9} && 5\\

\noalign{\smallskip}
\hline
\end{array}
$$
\noindent
Refs. --  1) \citet{Wong04}, 2) \citet{Lindqvist99}, 3) Hipparcos catalogue, 4) \citet{Kholopov85}, 5) \citet{Olofsson00}. 

\end{table}

\section{Observations}
\subsection{Molecular line observations}
\citet{Olofsson96} published molecular line spectra obtained towards all the stars
in Table~\ref{radio}, but \object{U~Cam} and \object{DR~Ser}.
Additional (sub)millimetre single-dish spectra used in the present analysis have been published in \citet{Olofsson93b, Olofsson93a}, \citet{Schoeier01}, and \citet{Wong04}. 

In this paper we present some complementary observational data. The  James Clerk Maxwell Telescope\footnote{The JCMT is operated by the Joint Astronomy Centre in Hilo, Hawaii on behalf of the
present organizations: the Particle Physics and Astronomy Research Council in the 
United Kingdom, the National Research Council of Canada and the Netherlands
Organization for Scientific Research.}  (JCMT) located at Mauna Kea was used to collect CO $J=3\rightarrow2$ (345.796 GHz) and $J=4\rightarrow3$ (461.041 GHz) line emission from   \object{R~Scl}, \object{DR~Ser}, and \object{U~Cam}. The observations were performed  in
 June, July, and October 2003, except for the $J=4\rightarrow3$ observations towards \object{U~Cam} which were done in November 2000.  CO $J=1\rightarrow0$ (115.271 GHz) and $J=2\rightarrow1$ (230.538 GHz) line emission  from   \object{U~Cam} was obtained in November 1999 using the IRAM 30\,m telescope\footnote{The IRAM 30\,m telescope is operated by the Intitut de Radio Astronomie Millim\'trique, which is supported by the Centre National de Recherche Scientifique (France), the Max Planck Gesellschaft (Germany) and the Instituto Geogr\'afico National (Spanin).} at Pico Veleta. In addition, CO  $J=1\rightarrow0$ line emission towards \object{TT Cyg} and  $^{13}$CO  $J=2\rightarrow1$ (220.399 GHz) line emission towards \object{V644~Sco} were detected using SEST in April 2002.  The HCN $J=1\rightarrow0$ (88.632 GHz) data for \object{DR~Ser} were collected during 1996-1997 using the Onsala Space Observatory (OSO) 20\,m telescope\footnote{The Onsala 20\,m telescope is operated by the  Swedish National Facility for Radio Astronomy, Onsala Space observatory at Chalmers University of technology}. HCN $J=1\rightarrow0$ and $J=3\rightarrow2$ (265.886 GHz) observations of \object{V644~Sco} were obtained using the Swedish-ESO submillimetre telescope\footnote{The SEST was operated jointly by the Swedish National Facility for Radio Astronomy and the European Southern Observatory (ESO).} (SEST) at La Silla in April 2002.  

\begin{table}
\caption[]{Summary of single-dish millimetre line observations used in the analysis.}
\label{intensities_summary}
$$ 
\begin{array}{p{0.25\linewidth}llcc}
\hline
\noalign{\smallskip}
\multicolumn{1}{l}{{\mathrm{Source}}} &
\multicolumn{1}{c}{{\mathrm{Telescope}}} &
\multicolumn{1}{c}{{\mathrm{Transition}}} &
\multicolumn{1}{c}{\int T_{\mathrm{mb}}\mathrm{d}v}  &
\multicolumn{1}{c}{{\mathrm{Ref.}}} 
\\
&
 &
 & 
\multicolumn{1}{c}{\mathrm{[K\,km\,s}^{-1}]}\\
\noalign{\smallskip}
\hline
\noalign{\smallskip}

R Scl         & \mathrm{IRAM} & \mathrm{CO(1-0)} &                      76.2 & 4\\
                   & \mathrm{IRAM} & \mathrm{CO(2-1)} &                      67.7 & 4\\
                   & \mathrm{SEST} & \mathrm{CO(1-0)} &                      26.1 & 4\\
                   & \mathrm{SEST} & \mathrm{CO(2-1)} &                      47.8 & 4\\
                   & \mathrm{SEST} & \mathrm{CO(3-2)} &                     63.7 & 3\\
                   & \mathrm{JCMT} & \mathrm{CO(2-1)} &                     50.6 & 6\\
                   & \mathrm{JCMT} & \mathrm{CO(3-2)} &                      62.8 & 1\\
                   &  \mathrm{JCMT} & \mathrm{CO(4-3)} &                      63.2 & 1\\
                   &  \mathrm{SEST} & ^{13}\mathrm{CO(1-0)} &  \phantom{0}1.9 & 5\\
                   &  \mathrm{SEST} & ^{13}\mathrm{CO(2-1)} & \phantom{0}4.2 & 5\\
                   &  \mathrm{SEST} & \mathrm{HCN(1-0)} & \phantom{0}1.3 & 3\\
                   &  \mathrm{SEST} & \mathrm{HCN(3-2)} & \phantom{0}7.7 & 3\\
                   &  \mathrm{SEST} & \mathrm{HCN(4-3)} & \phantom{0}3.9 & 3\\
                   &  \mathrm{HHT} & \mathrm{HCN(3-2)} & \phantom{0}2.4 & 7\\
                   &  \mathrm{HHT} & \mathrm{HCN(4-3)} & \phantom{0}4.1 & 7\\
                   
U Cam     & \mathrm{IRAM} & \mathrm{CO(1-0)} &                             21.8 &  1  \\
                  & \mathrm{IRAM} & \mathrm{CO(2-1)} &                             97.9 &  1  \\ 
                  & \mathrm{JCMT} & \mathrm{CO(3-2)} &                             66.7 & 1  \\
                  & \mathrm{JCMT} & \mathrm{CO(4-3)} &                              54.6 & 1 \\

U Ant        & \mathrm{SEST} & \mathrm{CO(1-0)} &                      10.8 &  3  \\
                  & \mathrm{SEST} & \mathrm{CO(2-1)} &                      13.1 &  3  \\ 
                  & \mathrm{SEST} & \mathrm{CO(3-2)} &   \phantom{0}9.7 & 3  \\

V644 Sco & \mathrm{SEST} & \mathrm{CO(1-0)} &          \phantom{0}7.6  & 3\\ 
                   & \mathrm{SEST} & \mathrm{CO(2-1)} &                             20.7  & 3\\ 
                   & \mathrm{SEST} & \mathrm{CO(3-2)} &                             21.4  & 3\\ 
                   & \mathrm{JCMT} & \mathrm{CO(2-1)} &                             22.2  & 6\\ 
                   & \mathrm{JCMT} & ^{13}\mathrm{CO(2-1)} & \phantom{0}1.1 &1\\
                   & \mathrm{SEST} & \mathrm{HCN(1-0)} &       \phantom{0}0.3 & 1 \\ 
                   &\mathrm{SEST}  & \mathrm{HCN(3-2)} &       \phantom{0}0.4 & 1 \\

DR Ser    & \mathrm{SEST} & \mathrm{CO(1-0)} &            \phantom{0}3.7 & 2 \\
                  & \mathrm{NRAO} & \mathrm{CO(2-1)} &                              11.3 & 2 \\
                  & \mathrm{JCMT} & \mathrm{CO(3-2)} &                               14.7 & 1 \\
                  &\mathrm{JCMT}  & \mathrm{CO(4-3)} &           \phantom{0}8.5 & 1 \\
                  & \mathrm{OSO} & \mathrm{HCN(1-0)} &          \phantom{0}0.6 & 1\\
                    
S Sct         & \mathrm{SEST} & \mathrm{CO(1-0)} &                      \phantom{0}5.8 &  3  \\
                  & \mathrm{SEST} & \mathrm{CO(2-1)} &                      \phantom{0}4.3 &  3  \\ 
                  & \mathrm{SEST} & \mathrm{CO(3-2)} &                      \phantom{0}3.1 & 3  \\
                  & \mathrm{SEST} & ^{13}\mathrm{CO(1-0)} &             \phantom{0}0.2 & 3 \\
                 
TT Cyg     & \mathrm{SEST} & \mathrm{CO(1-0)} &        \phantom{0}5.0 &  1  \\
                   & \mathrm{OSO} & \mathrm{CO(1-0)} &         \phantom{0}6.4 &  3  \\
                  & \mathrm{IRAM} & \mathrm{CO(1-0)} &         \phantom{0}7.5 &  3  \\ 
                  & \mathrm{IRAM} & \mathrm{CO(2-1)} &          \phantom{0}6.0 &  3  \\ 
                 
\noalign{\smallskip}
\hline
\end{array}
$$ 
\noindent
Refs. --  1) This paper. 2) \citet{Schoeier01}. 3) \citet{Olofsson96}. 4) \citet{Olofsson93a}. 5) \citet{Schoeier00}. 6) JCMT public archive. 7) \citet{Bieging01}.
\end{table}
\begin{figure*}
\centerline{\includegraphics[width=16cm]{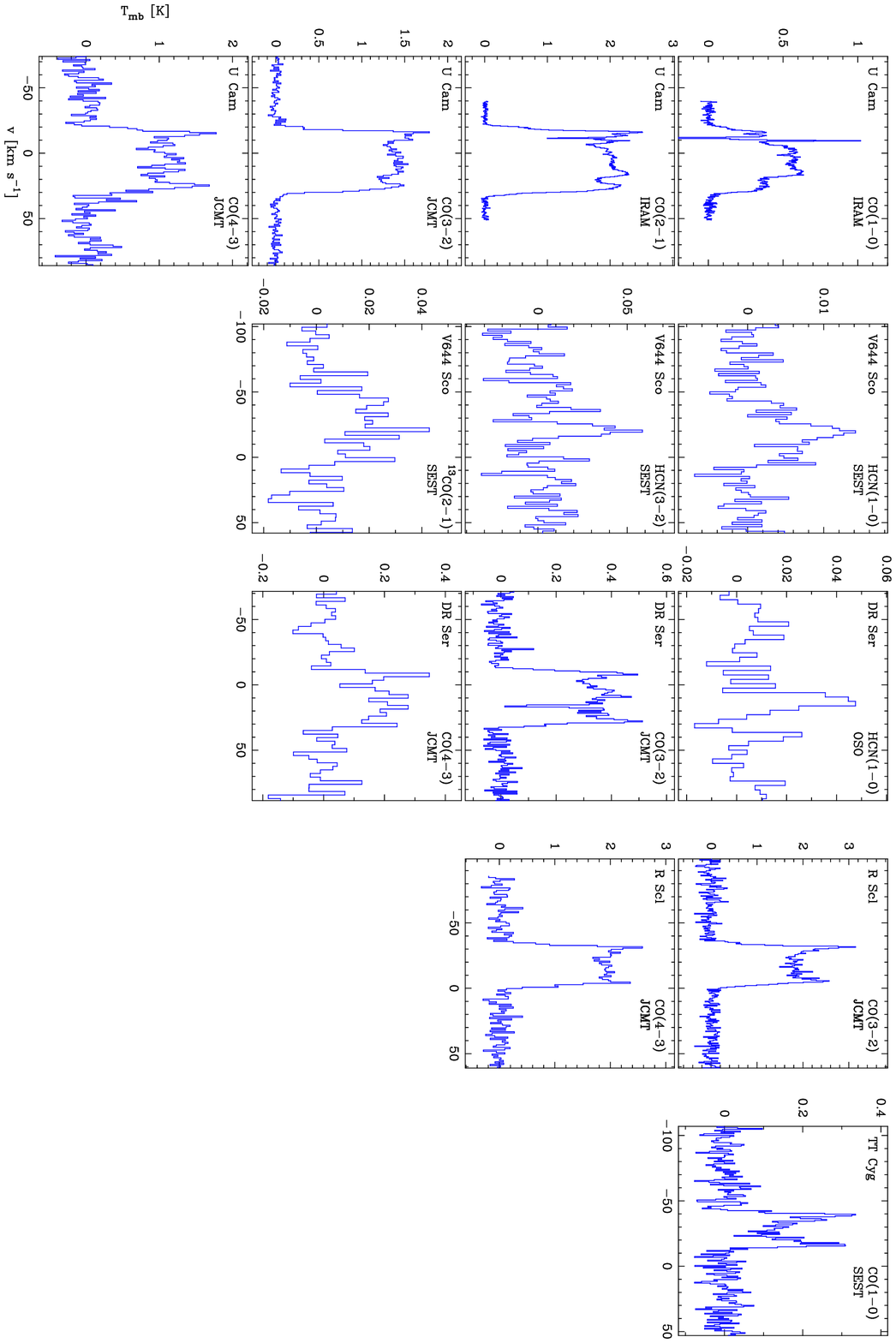}}
  \caption{New observations of CO and HCN radio line emission.} 
  \label{spectra}
\end{figure*}

All observations were made in a dual beamswitch mode, 
where the source is alternately placed in the signal and the reference
beam, using a beam throw of about $2\arcmin$ at JCMT,  $4\arcmin$ at  IRAM, $11\arcmin$ at OSO, and $12\arcmin$ at SEST. 
This method produces very flat baselines.  
Regular pointing checks made on strong continuum sources (JCMT and IRAM) 
or SiO masers (OSO and SEST) were found to be consistent within $\approx$\,3$\arcsec$. 

The observed spectra are presented in Fig.~\ref{spectra} and velocity-integrated intensities are reported in Table~\ref{intensities_summary}. The intensity
scales are given in main beam brightness temperature, $T_{\mathrm{mb}}$\,=\,$T_{\mathrm
A}^{*}/\eta_{\mathrm{mb}}$, where $T_{\mathrm A}^{\star}$ is the
antenna temperature corrected for atmospheric attenuation using the
chopper wheel method, and $\eta_{\mathrm{mb}}$ is the main beam
efficiency. For the CO $J=3\rightarrow2$ and $J=4\rightarrow3$ observations at the JCMT we have used main-beam efficiencies of 0.62 and 0.5, respectively. For the IRAM CO $J=1\rightarrow0$ and $J=2\rightarrow1$ observations we have used main-beam efficiencies ($B_{\mathrm{eff}}/F_{\mathrm{eff}}$) of 0.7 and 0.5, respectively. 
For the HCN $J=1\rightarrow0$ data obtained at OSO $\eta_{\mathrm{mb}} = 0.65$ was used. For the SEST  $\eta_{\mathrm{mb}} =0.75$ was used for the HCN $J=1\rightarrow0$ data, $\eta_{\mathrm{mb}} =0.7$ for the CO $J=1\rightarrow0$ data, and  $\eta_{\mathrm{mb}} =0.5$ for the HCN $J=3\rightarrow2$  and $^{13}$CO  $J=2\rightarrow1$ observations.
The uncertainty in the absolute intensity scale is estimated to be about $\pm 20$\%. 

The data was reduced in a standard way, by removing a first order baseline and then binned in order to improve the 
signal-to-noise ratio, using XS\footnote{XS is a package developed by P. Bergman to reduce and analyse a large number of single-dish spectra. It is publically available from {\tt ftp://yggdrasil.oso.chalmers.se}}.

A summary of the available data, relevant to the analysis in this paper, are given in Table~\ref{intensities_summary}.

\begin{figure*}
\centerline{\includegraphics[width=18cm]{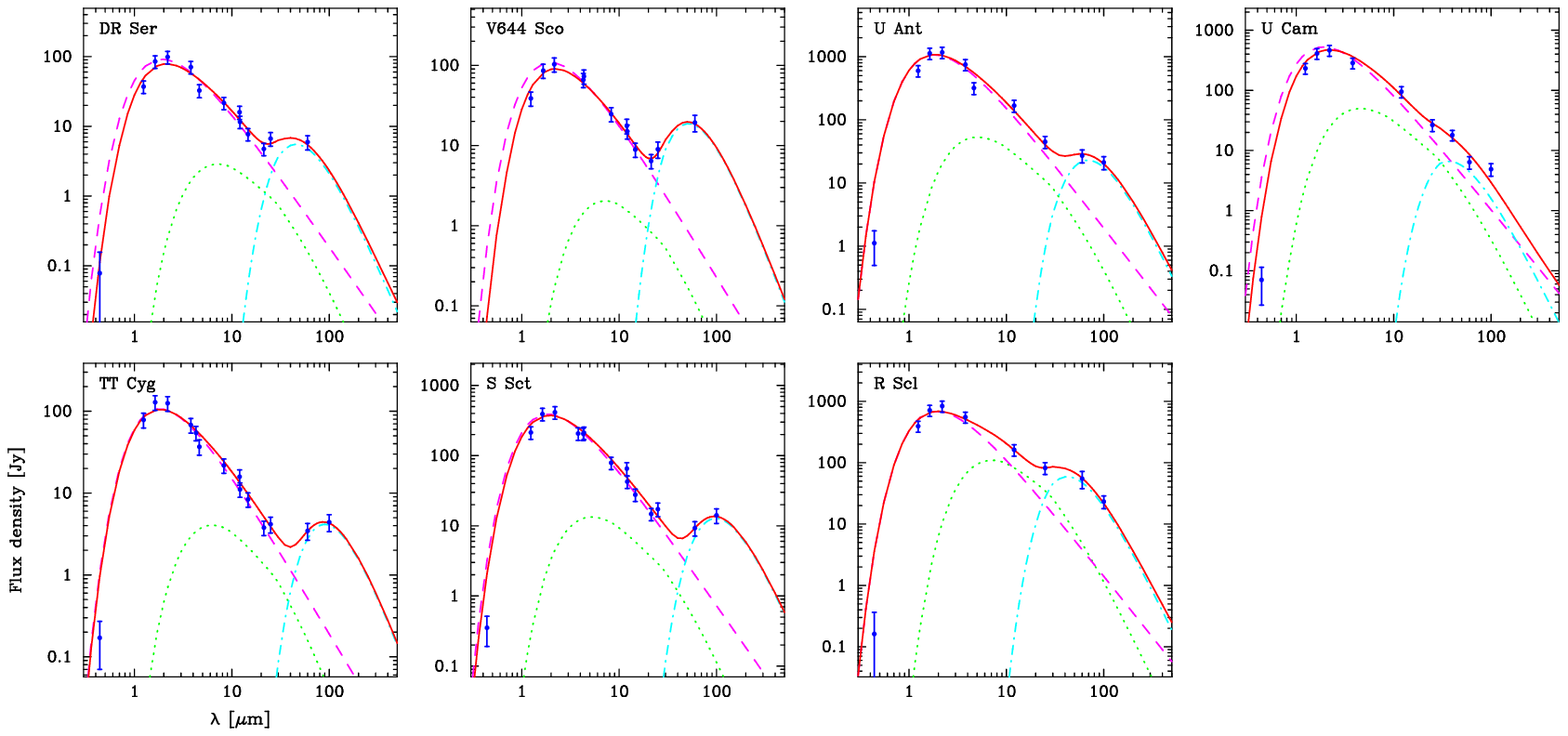}}
  \caption{Observed SEDs for the seven carbon stars with detected detached molecular shells. Best-fit models to the observed SEDs are overlayed (solid line). The model SEDs are composed of a stellar part (dashed line), a present-day wind (dotted line), and a detached shell (dash-dotted line). The model SEDs have been corrected for interstellar extinction. Stellar and envelope parameters are summarized in  Tables~\ref{stellar} and \ref{dust}.} 
  \label{SED}
\end{figure*}

\subsection{Continuum observations}
\label{obs_cont}
The observed spectral energy distribution of each source is used to constrain the dust modelling. We have selected observations that contain the total source flux at a specific wavelength. 
In addition to IRAS fluxes (12--100~$\mu$m), near-infrared (NIR) {\em JHKLM}-photometric data have been obtained from the literature (see Kerschbaum 1999\nocite{Kerschbaum99} and references therein). The IRAS
and NIR data for a particular source were generally not obtained at the same epoch, while the NIR photometry data
was. No reliable 100\,$\mu$m IRAS flux estimates are available for \object{DR~Ser}
 and \object{V644~Sco}, due to cirrus contamination. For \object{V644~Sco}, \object{DR~Ser}, \object{S~Sct}, and  \object{TT~Cyg}  there exist observations from the Midcourse Space Experiment (MSX) in the range $8-21 \mu$m which have been used in the analysis. 

\object{S~Sct} has the largest detached shell of the sample sources with a diameter of approximately $140\arcsec$. This is comparable to the size of the IRAS beam at 60~$\mu$m ($120\arcsec$) and resolution effects start to become important. \citet{Groenewegen94}, in their modelling of \object{S~Sct}, estimated that about 30\% of the flux was resolved out. Given the larger beam ($220\arcsec$) the effect at 100~$\mu$m is much lower as it is for all other sources in the sample. For this reason resolution effects has not been taken into account in the analysis. 

We have obtained archival data taken by the Short Wavelength Spectrometer (SWS) onboard the Infrared Space Observatory (ISO)  for \object{U~Cam} and estimated the flux at 
40~$\mu$m.  \object{U~Cam}  is the source with the smallest detached shell of $15\arcsec$ in diameter, smaller than the ISO aperture ($17\arcsec\times40\arcsec$) at 40~$\mu$m.


The observed SEDs are presented in Fig.~\ref{SED}.

\section{Radiative transfer}
This section describes the basic assumptions made in the analysis of the observed molecular line and continuum emission, and the methods used in the treatment of the radiative transfer.

\subsection{General considerations}
The CSEs are assumed to consist of two spherically symmetric components: an attached CSE (aCSE), i.e., a normal continuous AGB wind, and a detached, geometrically thin, CSE (dCSE) of constant density and temperature. Both components are assumed to be expanding at constant, but different, velocities. From the observed line spectra and SEDs  it is clear that in many cases it is not easy to separate the contribution from the present-day wind and the detached shell. In these cases the parameters describing the shell need to be varied together with the mass-loss rate describing the present-day wind. 

In what follows the procedures applied to determine the properties of the shells, as well as those of the present-day winds, from both continuum and  millimetre line observations, are descibed. 

The best fit model is found by minimizing  
\begin{equation}
\label{chi2_sum}
\chi^2 = \sum^N_{i=1} \left [ \frac{(F_{\mathrm{mod},\lambda}-F_{\mathrm{obs},\lambda})}{\sigma_{\lambda}}\right ]^2, 
\end{equation} 
where $F_{\lambda}$ is the flux, and $\sigma_{\lambda}$ the uncertainty in the measured 
flux, at wavelength $\lambda$, and the summation is done over
$N$ independent observations. 

\subsection{Molecular line modelling}
A detailed non-LTE radiative transfer code, based on the Monte Carlo method
\citep{Bernes79}, is used to perform the excitation analysis and to model the observed  
circumstellar line emission.  
The code is described in detail in \citet{Schoeier01} and has been
benchmarked, to high accuracy, against a wide variety of molecular line radiative
transfer codes in \citet{Zadelhoff02} and van der Tak et al.\ (in prep.).  

The excitation analysis  includes 
radiative excitation through vibrationally excited states, and a full
treatment of line overlaps between various hyperfine components in  the case of HCN,
see e.g. \citet{Lindqvist00}. Relevant molecular data are summarized in \citet{Schoeier05a} and are made publicly available at {\tt http://www.strw.leidenuniv.nl/$\sim$moldata}.

\begin{table}
\caption[]{Stellar properties.}
\label{stellar}
$$ 
\begin{array}{p{0.25\linewidth}ccccccccc}
\hline
\noalign{\smallskip}
\multicolumn{1}{l}{\mathrm{Source}} &
\multicolumn{1}{c}{\mathrm{Var.\ type}} &&
\multicolumn{1}{c}{P} &&
\multicolumn{1}{c}{D} &&
\multicolumn{1}{c}{L_{\star}} &&
\multicolumn{1}{c}{T_{\star}}\\
&
&&
\multicolumn{1}{c}{\mathrm{[days]}} &&
\multicolumn{1}{c}{[\mathrm{pc}]} &&
\multicolumn{1}{c}{[\mathrm{L}_{\sun}]} &&
\multicolumn{1}{c}{[\mathrm{K}]} 
\\
\noalign{\smallskip}
\hline
\noalign{\smallskip}
R Scl        & \mathrm{SRb}&& 370 && 290^2 && 4300 && 2625 \\
U Cam      &  \mathrm{SRb} &&  400 && 430^2 &&7000 && 2695 \\
U Ant        &  \mathrm{Lb}  && \mbox{---} && 260^1 && 5800 && 2810 \\
V644 Sco &  \mathrm{Lb}  && \mbox{---} && 700^3 && 4000 && 2615 \\
DR Ser     &  \mathrm{Lb}  &&\mbox{---} && 760^3 && 4000 && 2650 \\
S Sct         &  \mathrm{SRb} && 148 && 400^1 && 4900 && 2755\\
TT Cyg     &  \mathrm{SRb} && 118  && 610^1 && 3200 && 2825\\
\noalign{\smallskip}
\hline
\end{array}
$$ 
$^1$ Hipparcos distance \citep{Knapp03}\\
$^2$ From period-luminosity relation \citep{Knapp03}\\
$^3$ From adopting $L_{\star}$\,=\,4000\,L$_{\odot}$\\

\end{table}

\subsection{Continuum modelling}
An independent estimate of the properties of the detached shell and the present-day
wind can be made from the  excess emission observed at wavelengths longer than 
$\approx$\,5\,$\mu$m and 20\,$\mu$m, respectively. The dust emissions from both the shell and the present-day wind are expected to be optically thin at all relevant wavelengths ($\gtrsim$\,1\,$\mu$m). The assumption of optically thin emission greatly simplifies the radiative transfer of the continuum emission. The simple model presented in detail below has been verified against a full radiative transfer code (Sch\"oier, in prep.) for the densest winds encountered in this project, and they are found to agree within the numerical accuracy of the code itself ($\lesssim 5\%$).

In the optically thin limit the contribution to the SED can be separated into a stellar component $F^{\mathrm{star}}_{\nu}$, a component from the present-day wind $F_\nu^{{\mathrm{wind}}}$, and
that from the detached shell $F_\nu^{{\mathrm{shell}}}$, which are solved for individually. The total SED is then obtained from $F_{\nu} = F^{\mathrm{star}}_{\nu} +  F_\nu^{{\mathrm{wind}}} + F_\nu^{{\mathrm{shell}}}$ (note that the violation of energy conservation introduced in this way is negligible, $\lesssim$\,5\%). Note that we ignore any beam filling effects (see Sect.~\ref{obs_cont}).
 
The flux $F_\nu^{\mathrm{wind}}$ received by an observer located at a distance $D$ from an expanding, spherically symmetric wind produced by a constant rate of mass-loss is
\begin{equation}
\label{Fwind}
F_\nu^{{\mathrm{wind}}} = \frac{4\pi^2a^2_{\mathrm{gr}}Q_{\nu}}{D^2} \int^{R_{\mathrm{e}}}_{R_{\mathrm{i}}} n_{\mathrm{dust}}(r)B_{\nu}(T_{\mathrm{dust}}(r)) r^2 \mathrm{d}r,
\end{equation}
where $Q_{\nu}$ is the absorption efficiency of a spherical dust grain with radius $a_{\mathrm{gr}}$, $n_{\mathrm{dust}}$  the number density of dust grains, $B_{\nu}$ the blackbody brightness at the dust temperature $T_{\mathrm{dust}}$, and 
the integration is made from the inner ($R_{\mathrm i}$) 
to the outer ($R_{\mathrm e}$) radius of  the CSE. 

For an extended, geometrically thin, isothermal shell at constant density, adequate for describing the detached shells treated here,  Eq.~(\ref{Fwind}) reduces to
\begin{equation}
F_\nu^{{\mathrm{shell}}} = \frac{9 Q_{\nu}}{4 a_{\mathrm gr} \rho_{\mathrm{gr}}D^2} B_{\nu}(T_{\mathrm{dust}}) M_{\mathrm{dust}},
\end{equation}
where the dust mass of the shell, $M_{\mathrm{dust}}$, and the specific density
of a grain, $\rho_{\rm gr}$, have been introduced. 

The local dust temperature, $T_{\mathrm{dust}}(r)$, is determined by balancing the heating ($H$) provided by absorption of infalling radiation and cooling ($C$) by subsequent re-emission. The heating term can be expressed as
\begin{equation}
\label{heating}
H = \pi a^2_{\mathrm{gr}} n_{\mathrm{dust}}(r) \int F^{\mathrm{star}}_{\nu}(r) Q_{\nu} \mathrm{d}\nu,
\end{equation}
where the dust grains are assumed to be of the same size and spherical. The dust grains are assumed to locally radiate thermal emission according to Kirchoff's law and
the resulting cooling term is 
\begin{equation}
\label{cooling}
C = 4\pi^2 a^2_{\mathrm{gr}} n_{\mathrm{dust}}(r) \int B_{\nu}(T_{\mathrm{dust}}(r)) Q_{\nu} \mathrm{d}\nu.
\end{equation}

Modelling the SED provides information only on the density structure of the dust grains, $\propto \dot{M}_{\mathrm{dust}}/v_{\mathrm{dust}}$, where $\dot{M}_{\mathrm{d}}$ is the dust-mass-loss rate and $v_{\mathrm{dust}}$ the dust velocity.
In order to calculate $\dot{M}_{\mathrm{dust}}$, and eventually the gas mass-loss rate $\dot{M}_{\mathrm{gas}}$, one has to rely upon some model describing the dynamics of the wind. There are now growing evidence that the winds of AGB stars are driven by radiation pressure from stellar photons exerted on dust grains. Through momentum-coupling between the dust and gas, molecules are dragged along by the grains. The stationary solution to such a problem has been extensively studied \citep[e.g.,][]{Habing94}. The terminal (at large radii) gas velocity, $v_{\mathrm{gas}}$, is given by
\begin{equation}
v_{\mathrm{gas}} = \sqrt{2GM_{\star}(\Gamma - 1)/r_0},
\end{equation}
where $G$ is the gravitational constant, $M_{\star}$ the stellar mass, $L_{\star}$ the luminosity, and $\Gamma$ the ratio of the drag force and the gravitational  force on the gas, given by
\begin{equation}
\Gamma = \frac{3(Q_{\mathrm F}/a_{\mathrm{gr}})L_{\star}\Psi}{16\pi GcM\rho_{\mathrm{gr}}}\frac{v_{\mathrm{gas}}}{v_{\mathrm{dust}}}, 
\end{equation}
where $Q_{\mathrm F}$ is the flux-averaged momentum-transfer efficiency, $c$ the speed of light, and $\Psi$ the dust-to-gas mass-loss-rate ratio. The drift velocity $v_{\mathrm{dr}} = v_{\mathrm{dust}} - v_{\mathrm{gas}}$ can be obtained from
\begin{equation}
v_{\mathrm{dr}} = \sqrt{\frac{v_{\mathrm{gas}}Q_{\mathrm F} L_{\star}}{c\dot{M}_{\mathrm{gas}}}}.
\end{equation}

In the modelling,  amorphous carbon dust grains with the optical constants presented in 
\citet{Suh00} were adopted. The properties of these grains were shown to reproduce,
reasonably well, the SEDs from a sample of carbon stars. 
For simplicity, the dust grains are assumed to be spherical and of the same size with $a_{\mathrm{gr}}$\,=\,0.1\,$\mu$m and $\rho_{\rm gr}$\,=\,2\,g\,cm$^{-3}$. The absorption and scattering efficiencies were then calculated using standard Mie theory \citep{Bohren83}. For the dynamical solution we further assume a stellar mass of $M$\,=\,1\,M$_{\odot}$ and adopt an inner radius of the envelope $r_0$\,=\,8$\times$10$^{13}$\,cm (corresponding to a dust condensation temperature of $\approx$\,1200\,K).

\section{Dust modelling}
\label{dust_modelling}
The state of the dust grains will have a great impact on the gas modelling, affecting both its dynamics and molecular excitation. However, the reverse is not true so the dust modelling can safely be done without bothering about the state of the gas, except for the terminal gas velocity  which enters in the solution of the dynamics and allows the total mass-loss rate to be estimated. Fortunately, the terminal gas velocity is an observable.

\subsection{Stellar parameters}
The first step in the modelling is to determine the luminosity (alternatively, the distance) from fitting the SED shortward of $\approx$\,10\,$\mu$m where the stellar emission fully dominates in these optically thin envelopes. The stellar emission, $F_{\nu}^{\mathrm{star}}$, is assumed to be that of a blackbody. The stellar effective temperatures, $T_{\star}$, for our sample sources were adopted from \citet{Bergeat01}. Accurate Hipparcos parallaxes exist only for \object{U~Ant}, \object{S~Sct}, and \object{TT~Cyg} \citep{Knapp03}, and other means of determining the distance to the remaining objects need to be adopted. For \object{R~Scl} and \object{U~Cam} we used the period-luminosity relation by \citet{Knapp03} to obtain the distances, adopting the periods listed by \citet{Kholopov85} and \citet{Knapp03}. The distances to the irregular variables  \object{V644~Sco} and \object{DR~Ser} were obtained after adopting a luminosity of 4000~L$_{\odot}$. All the stellar parameters are listed in Table~\ref{stellar}. In the fitting procedure, interstellar extinction was accounted for by adopting the extinction law of  \citet{Cardelli89} and using the vales for extinction in the visual, $A_{\mathrm V}$, listed in \citet{Bergeat01}. 

\subsection{Detached shell and present-day wind}
The observational constraints, in the form of SEDs covering the wavelength range 
$\approx$\,10\,$-$\,100\,$\mu$m are analyzed using the $\chi^2$-statistic defined in Eq.~(\ref{chi2_sum}).
A calibration uncertainty of 20\% was added to 
the total error budget and will, in most cases, dominate the error which in turn means that
fluxes at various wavelengths typically have the same weight in the $\chi^2$-analysis.
It should be noted that the present-day mass loss is more difficult to estimate than the properties of the detached shell given its lower contrast with respect to the stellar contribution.

\begin{figure}
\centerline{\includegraphics[width=7cm]{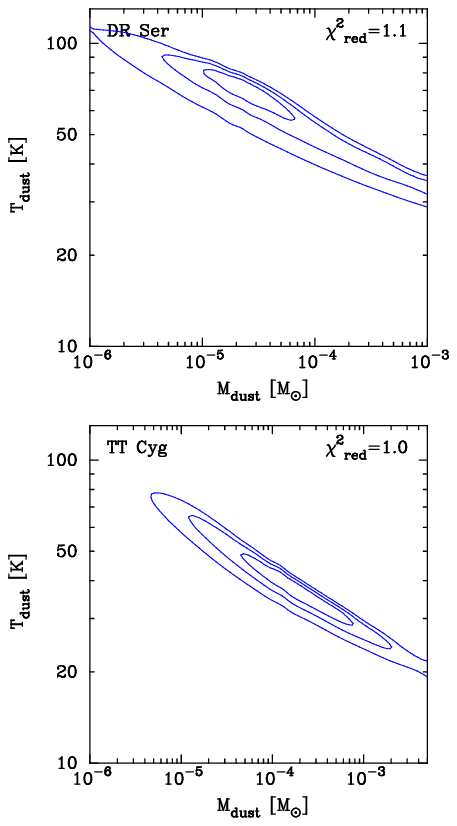}}
  \caption{$\chi^2$-maps showing the sensitivity of the adjustable parameters  $T_{\mathrm{dust}}$ and $M_{\mathrm{dust}}$ in the dust modelling of the detached shells around \object{DR~Ser} and \object{TT~Cyg}. Contours are drawn at $\chi^2_{\mathrm{min}} + (2.3, 6.2, 11.8)$ indicating the 68\% (`1$\sigma$'), 95\%
(`2$\sigma$'), and 99.7\% (`3$\sigma$') confidence levels,
respectively.  The quality of the best fit model can be
estimated from the reduced chi-squared statistic $\chi^{2}_{\mathrm{red}}$=$\chi^{2}_{\mathrm{min}}$/($N$$-$2)
shown in the upper right corner.} 
  \label{dust_chi2}
\end{figure}

For the present-day wind there is only one adjustable parameter in the modelling, the gas-mass-loss rate. The emission from the shell is fully described by adjusting the temperature of the dust grains ($T_{\mathrm{dust}}$) and the total dust mass in the shell ($M_{\mathrm{dust}}$). The best fit models all have reduced $\chi^2$ values of $\approx 1$, indicating good fits. The parameters obtained are listed in Table~\ref{dust}. The dust-to-gas ratio ($\Psi$) in the present-day wind obtained from the dynamical model is also listed. The sensitivity of the results to the adjustable parameters are illustrated in Fig.~\ref{dust_chi2} for the young shell around \object{DR~Ser} and the older shell around \object{TT~Cyg}. The final fits to the observed SEDs are presented in Fig.~\ref{SED}, where also the contribution of each of the three components (stellar, present-day wind, and detached shell) are indicated.

\begin{figure}
\centerline{\includegraphics[width=7cm]{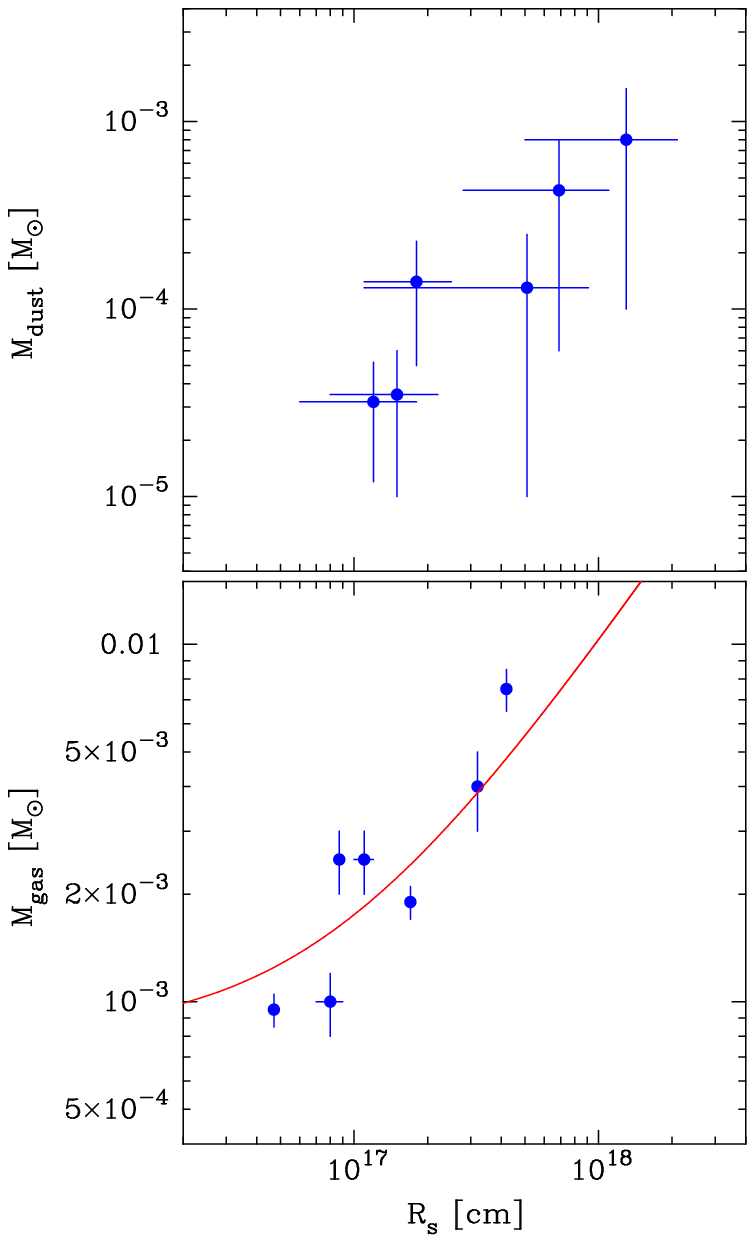}}
  \caption{The shell dust ($M_{\mathrm{dust}}$; upper panel) and gas ($M_{\mathrm{gas}}$; lower panel) masses plotted as a function of the radius of the shell ($R_{\mathrm s}$).  The solid line shows the mass-radius relation of the interacting wind model presented in Sect.~\ref{interaction_model}.} 
  \label{mass_radius}
\end{figure}

The dust temperature in the shell can be translated into a radial distance of the shell from balancing the heating [Eq.~(\ref{heating})], provided by the central star, and the cooling [Eq.~(\ref{cooling})],  provided by the re-emission of the observed photons at longer wavelengths, as in the case of the present-day wind. The distances obtained are given in Table~\ref{dust}. As shown in Fig.~\ref{mass_radius}, although the errors involved are quite large, there is a clear correlation between the radial distance of the shell and its mass, $M_{\mathrm{dust}} \propto R_{\mathrm{s}}^{1.36\pm0.53}$. Pearson's correlation coefficient $r$ is 0.97 indicating an almost complete positive correlation. The fact that the mass increases with distance suggests that matter is being swept up as the shell expands away from the star. Such a scenario will be further elaborated in Sect.~\ref{interaction}.

\object{U~Cam} is the source which has the lowest contrast between the stellar and shell contributions to the SED.  It is the youngest shell in the sample located close to the central star,  Sect.~\ref{mol}., complicating also its separation from the present day wind. This makes it hard from the dust modelling to constrain the properties of the circumstellar dust for this particular source.

\citet{Groenewegen94} modelled the SED of \object{S~Sct} and found a shell dust mass of $\approx$\,2$\times$10$^{-4}$~M$_{\odot}$. Given the uncertainties in the adopted dust properties, this is consistent with the mass derived in the present analysis. Placing the dust shell at the same location as that of CO we derive a dust mass of $\approx$\,8$\times$10$^{-5}$~M$_{\odot}$ for our somewhat smaller distance.

\citet{Izumiura97} in their analysis of  HIRAS images found evidence for two separate dust shells  around \object{U~Ant} where the inner of the two shells is coinciding with that of CO. They derive a total mass for this shell of $\approx$\,5$\times$10$^{-3}$~M$_{\odot}$ consistent within the uncertainties with our estimate of $\approx$\,2$\times$10$^{-3}$~M$_{\odot}$ from the CO modelling. The outer shell, which has no known molecular counterpart, is similarly thought to have been produced in another thermal pulse some $\sim$\,10$^4$\,yr before. The dust mass contained in this shell is estimated to be $\approx$\,4$\times$10$^{-5}$~M$_{\odot}$ and its contribution to the SED is $\approx$\,15\% at 60\,$\mu$m.

\begin{table*}
\caption[]{Results from  dust modelling  (with 1$\sigma$ errors)}.
\label{dust}
$$ 
\begin{array}{p{0.15\linewidth}cccccccccccc}
\hline
\noalign{\smallskip}
&
\multicolumn{5}{c}{\mathrm{dCSE}} & &&&
\multicolumn{3}{c}{\mathrm{aCSE}} \\
\cline{2-6}\cline{10-12}
\multicolumn{1}{l}{\mathrm{Source}} &
\multicolumn{1}{c}{M_{\mathrm{dust}}} &&
\multicolumn{1}{c}{T_{\mathrm{dust}}}&&
\multicolumn{1}{c}{R_{\mathrm{s}}}&&&&
\multicolumn{1}{c}{\dot{M}_{\mathrm{gas}}} &&
\multicolumn{1}{c}{\Psi}  \\
&
\multicolumn{1}{c}{[\mathrm{M}_{\sun}]} &&
\multicolumn{1}{c}{[\mathrm{K}]} &&
\multicolumn{1}{c}{[\mathrm{cm}]}   &&&&
\multicolumn{1}{c}{[\mathrm{M}_{\sun}\,{\mathrm{yr}}^{-1}]} &&\\
\noalign{\smallskip}
\hline
\noalign{\smallskip}
R Scl        & (3.2\pm2.0)\times10^{-5} && 75\pm15 && (1.2\pm0.6)\times10^{17}  &&&&{\phantom{<}}3.8\times10^{-7} && 1.7\times10^{-3} \\
U Cam      & 1.5\times10^{-6} - 1.0\times10^{-3}    && 70\pm45 &&\mbox{---} &&&& {\phantom{<}}4.0\times10^{-7}  && 1.2\times10^{-3}   \\
U Ant        & (1.3\pm1.2)\times10^{-4} && 55\pm20 && (5.1\pm4.0)\times10^{17} &&&& {\phantom{<}}7.0\times10^{-8} && 3.0\times10^{-3} \\
V644 Sco & (1.4\pm0.9)\times10^{-4} && 61\pm9\phantom{0}   && (1.8\pm0.7)\times10^{17} &&&& <5.0\times10^{-8} && \mbox{---}\\
DR Ser     & (3.5\pm2.5)\times10^{-5} && 68\pm13 && (1.5\pm0.7)\times10^{17}&&&& <7.0\times10^{-8} &&   \mbox{---}\\
S Sct         & (8.0\pm7.0)\times10^{-4}  && 34\pm9\phantom{0} && (1.3\pm0.8)\times10^{18}&&&& <8.0\times10^{-8} && \mbox{---} \\
TT Cyg     & (4.3\pm3.7)\times10^{-4} && 39\pm10 && (6.9\pm4.1)\times10^{17} &&&& <3.2\times10^{-8} && \mbox{---}\\
\noalign{\smallskip}
\hline
\end{array}
$$ 
\smallskip
\end{table*}

\section{Molecular line modelling}
\label{mol}
\subsection{Line profiles}
Many of the observed CO spectra exhibit a  triple-peaked line profile, e.g., \citet{Olofsson96}. The, usually narrow, peaks at the extreme velocities arise from a spatially resolved shell, whereas the central component is produced by the present-day wind, i.e., a normal 
low-mass-loss-rate AGB wind.  Since the present-day wind is in general expanding at a significantly lower velocity, $v_{\rm p}$, than that of the shell, $V_{\rm s}$, emission at velocities between $v =v_\star \pm v_{\rm p}$  and  $v =v_\star \pm V_{\rm s}$ should contain only the contribution from the shell. Each of these velocity intervals is further divided into two separate components so that the total $\chi^2$-statistic also contains information on the line profile.

Information on the properties of the present-day mass-loss rate is contained in the intensity in the range $v=v_\star \pm v_{\rm p}$. However, there is also generally a contribution here from the detached shell and a good model for the shell is required in order to investigate the present-day wind. This makes it particularly difficult to 
separate the two components in all cases except \object{U~Ant}, \object{S~Sct}, and
\object{TT~Cyg}, \citep[e.g.,][]{Olofsson96}. In the case of \object{U~Cam} the
present-day wind is separated in the CO line interferometer data \citep{Lindqvist99}.
In the cases of \object{R~Scl}, \object{V644~Sco}, and \object{DR~Ser} the
detected HCN emissions indicate expansion velocities about a factor of two lower than in the case of the CO emission. The HCN emission is interpreted as arising predominantly in the present-day wind. This was also the conclusion reached by \citet{Wong04} for \object{R~Scl}. 

\begin{table*}
\caption[]{Results from molecular line modelling  (with 1$\sigma$ errors bars)}.
\label{gas}
$$ 
\begin{array}{p{0.15\linewidth}ccccccccccccc}
\hline
\noalign{\smallskip}
&
\multicolumn{7}{c}{\mathrm{dCSE}} & &&&
\multicolumn{3}{c}{\mathrm{aCSE}} \\
\cline{2-8}\cline{12-14}
\multicolumn{1}{l}{\mathrm{Source}} &
\multicolumn{1}{c}{M_{\mathrm{gas}}}& &
\multicolumn{1}{c}{T_{\mathrm{kin}}}&&
\multicolumn{1}{c}{R_{\mathrm{s}}}&
\multicolumn{1}{c}{V_{\mathrm{s}}} & 
\multicolumn{1}{c}{\mathrm{age}} & &&&
\multicolumn{1}{c}{\dot{M}_{\mathrm{gas}}} &
\multicolumn{1}{c}{v_{\mathrm{p}}} &
\multicolumn{1}{c}{\mathrm{Molecule}^1} \\
&
\multicolumn{1}{c}{[\mathrm{M}_{\sun}]}& &
\multicolumn{1}{c}{[\mathrm{K}]} &&
\multicolumn{1}{c}{[\mathrm{cm}]} &
\multicolumn{1}{c}{[\mathrm{km\,s}^{-1}]} &
\multicolumn{1}{c}{[\mathrm{yr}]} &&&&
\multicolumn{1}{c}{[\mathrm{M}_{\sun}\,{\mathrm{yr}}^{-1}]} &
 \multicolumn{1}{c}{[{\mathrm{km}}\,{\mathrm{s}}^{-1}]} &
\\
\noalign{\smallskip}
\hline
\noalign{\smallskip}
R Scl        & (2.5\pm0.5)\times10^{-3} && > 20\phantom{0} && 8.7\times10^{16} & 15.5& 1800 &&&&   3.0\times10^{-7} &  10.5 & \mathrm{HCN}  \\
U Cam      & (9.5\pm1.0)\times10^{-4} && >130 && 4.7 \times10^{16} & 23.0& \phantom{0}650 &&&& 2.0\times10^{-7} &  12.0 & \mathrm{CO} \\
U Ant        & (1.9\pm0.2)\times10^{-3} && >200 && 1.7\times10^{17} & 19.0 & 2800 &&&&  2.0\times10^{-8} & \phantom{0}4.0  & \mathrm{CO}  \\
V644 Sco & (2.5\pm0.3)\times10^{-3} && 170\pm100 && (1.1\pm0.2)\times10^{17} & 23.0& 1500 &&&& 5.0\times10^{-8} & \phantom{0}5.0  &  \mathrm{HCN} \\
DR Ser     & (1.0\pm0.2)\times10^{-3} && >100 && (8.0\pm1.5)\times10^{16} & 20.0& 1300 &&&& 3.0\times10^{-8}  &   \phantom{0}5.0      & \mathrm{HCN}\\
S Sct         & (7.5\pm1.0)\times10^{-3} && > 60\phantom{0} && 4.2\times10^{17} & 16.5& 8100 &&&& 2.0\times10^{-8} &  \phantom{0}4.0 & \mathrm{CO} \\
TT Cyg     & (4.0\pm1.0)\times10^{-3} && >200 && 3.2\times10^{17} & 12.5 & 8100 &&&& 3.2\times10^{-8} & \phantom{0}4.0  & \mathrm{CO} \\
\noalign{\smallskip}
\hline
\end{array}
$$ 
$^1$ The molecular line emission used for estimating the properties of the aCSEs. Note that in the analysis of the dCSEs CO line emission was used in all cases.
\end{table*}

\subsection{The detached shells}
In the analysis there exist four adjustable parameters: the H$_2$ number density ($n_{\mathrm{H}_2}$), kinetic gas temperature ($T_{\mathrm{kin}}$), thickness ($\Delta R_{\mathrm{s}}$) and location of the shell ($R_{\mathrm{s}}$). The fractional abundance of CO (in relation to H$_2$), $f_{\mathrm{CO}}$, is assumed to be equal to 1.0$\times$10$^{-3}$. 
In three cases (TT Cyg,
\citet{Olofsson00}; U Cam, \citet{Lindqvist99}; S Sct, Olofsson et al., in prep.) the sizes of the shells are well known from interferometric CO observations. In  the case of U Ant  the size of the shell is reasonably well known from single-dish CO mapping. \object{R~Scl} has a prominent detached shell at $\approx$\,8.7$\times$10$^{16}$\,cm seen in scattered light \citep{Delgado01}. 
In these five cases the position of the shell was fixed to match the observations. Although high-spatial-resolution observations exist, little is known about the actual  thickness of these shells. CO interferometric observations do not resolve the radial thickness of the shells. Here we adopt a value of 1.0$\times$10$^{16}$\,cm. The best-fit model is found from minimizing the $\chi^2$-statistic as described above.

\begin{figure}
\centerline{\includegraphics[width=7cm]{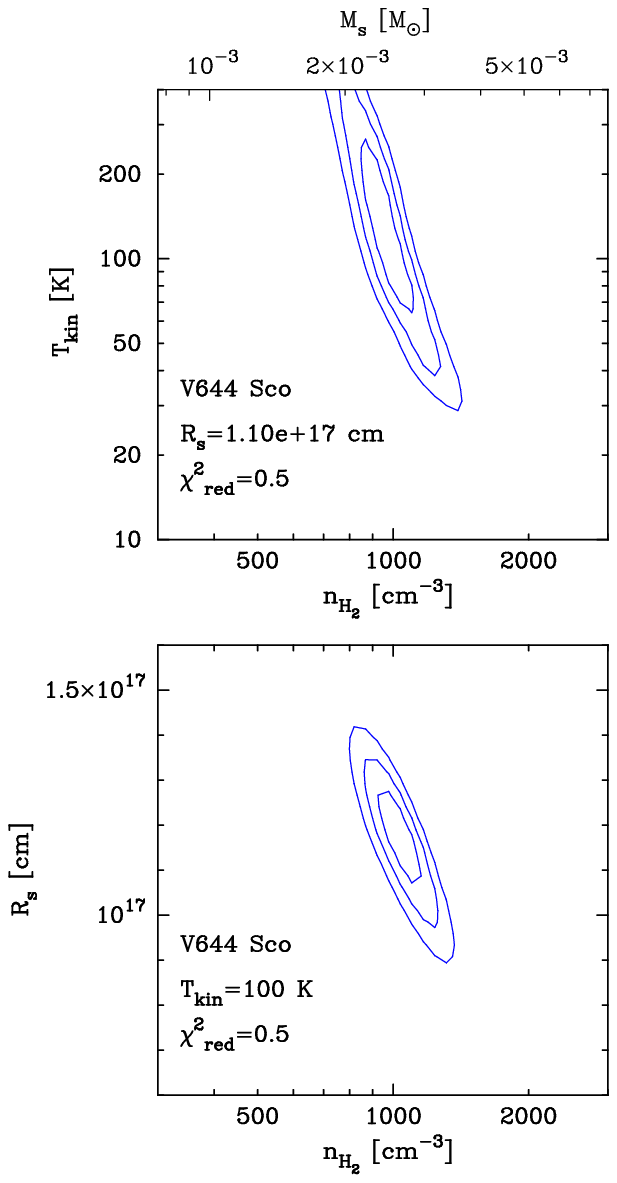}}
  \caption{$\chi^2$-maps showing the sensitivity of the adjustable parameters $n_{\mathrm{H_2}}$, $T_{\mathrm{kin}}$, and $R_{\mathrm s}$ in the CO modelling of the detached shell around \object{V644~Sco}. Contours are drawn at $\chi^2_{\mathrm{min}} + (2.3, 6.2, 11.8)$ indicating the 68\% (`1$\sigma$'), 95\%
(`2$\sigma$'), and 99.7\% (`3$\sigma$') confidence levels,
respectively.  The quality of the best fit model can be
estimated from the reduced chi-squared statistic $\chi^{2}_{\mathrm{red}}$=$\chi^{2}_{\mathrm{min}}$/($N$$-$2)
shown in the lower left corner. Also indicated is the value of the parameters that is kept fixed in these maps.} 
  \label{v644sco}
\end{figure}

As an example, in Fig.~\ref{v644sco} $\chi^2$-maps obtained from the modelling of the shell around \object{V644~Sco} are shown. The location of the shell ($R_{\mathrm{s}}$) is unknown and used as an adjustable parameter.  However, it turns out that the shell radial distance from the central star is the best constrained parameter in the modelling, together with the mass contained in the shell. Fig.~\ref{v644sco} (lower panel) shows the sensitivity of the model to changes in  $R_{\mathrm{s}}$ and $n_{\mathrm{H_2}}$ for a fixed temperature of 100\,K. In a broad temperature range we find that good fits are obtained for  $R_{\mathrm{s}}$=(1.1$\pm$0.2)$\times$10$^{17}$\,cm. Fixing the radial distance to 1.1$\times$10$^{17}$\,cm, the density and temperature are 
constrained to $1050\pm150$~cm$^{-3}$ and $170\pm100$~K, respectively (Fig.~\ref{v644sco}; upper panel). The mass contained in the shell is $M_{\mathrm{gas}}$\,=\,(2.5$\pm$0.3)$\times$10$^{-3}$\,M$_\odot$.

The derived mass in the shell is not very sensitive to the adopted thickness of the shell since the emission in almost all observed lines is optically thin. The density in the shell, however, goes roughly as $n_{\mathrm{H_2}}$\,$\propto$\,$\Delta R_{\mathrm{s}}^{-1}$. There is also a weak temperature dependence.  

The CO line emission from the lower rotational transitions is not very sensitive to changes in the kinetic gas temperature once $T_{\mathrm{kin}}$\,$>$\,50\,$-$\,100\,K. In order to better constrain the temperature in the shell transitions involving higher $J$-levels needs to be observed as illustrated in Sect.~\ref{sec_pred}. 

The properties of all dCSEs determined from the CO line modelling are summarized in Table~\ref{gas}. Fig.~\ref{mass_radius} illustrates that, as in  the case of the dust emission, there is a clear trend that the gas mass of the shell increases with its radial distance from the central star. Again a strong correlation is found $M_{\mathrm{gas}} \propto R_{\mathrm{s}}^{0.85\pm0.18}$ ($r=0.93$) further supporting a scenario where matter is being swept up as discussed in Sect.~\ref{interaction}. Also in indicated in Table~\ref{gas} is the age of the dCSEs estimated from $R_{\mathrm{s}}/V_{\mathrm{s}}$. Since there is a clear trend that $V_{\mathrm{s}}$ decrease with $R_{\mathrm{s}}$ (Sect.~\ref{interaction}) these ages are strictly upper limits to the age since the expansion was likely faster at earlier epochs.

\begin{figure*}
\centerline{\includegraphics[width=17cm]{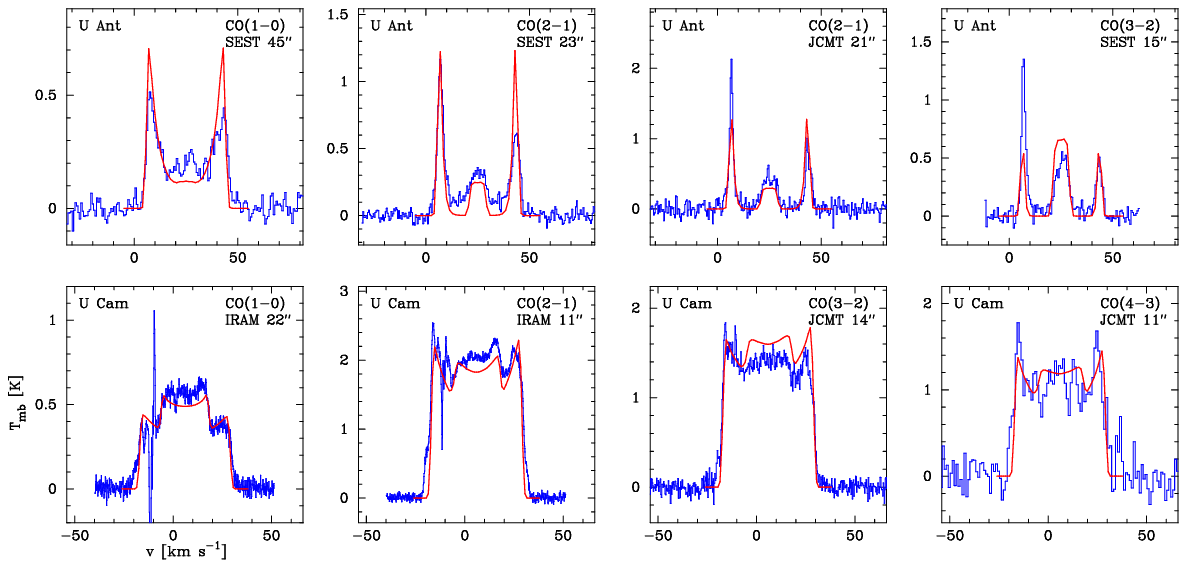}}
  \caption{Best fit models (solid line) including both the present-day wind and detached shell emission overlayed on observed spectra (histograms) for \object{U~Ant} and \object{U~Cam}. The uncertainty in the intensity scale is $\pm 20$\%. The narrow spikes found near the blue wing in the spectra of U~Cam are of interstellar origin. } 
  \label{models}
\end{figure*}

\subsection{The present-day winds}
Once the properties of the detached shell are determined and its contribution to the observed CO line emission is estimated, the corresponding estimates for the present-day wind can be can be made. In the modelling of the present-day wind the temperature structure is solved for self-consistently as described in \citet{Schoeier01} using the results from the dust modelling performed in Sect.~\ref{dust_modelling} to describe the properties of the dust entering in the heating of the gas envelope. It should be emphasised  that the actual CO line intensities are not very sensitive to the temperature structure since the excitation is dominated by the radiation field from the central star in these low-mass-loss-rate objects \citep{Schoeier01}.

Adopting a CO abundance of 1.0$\times$10$^{-3}$ relative to H$_2$ leaves just the mass-loss rate, $\dot{M}_{\mathrm{gas}}$, as an adjustable parameter  in the modelling. The expansion velocity of the present-day wind is fixed from matching the observed lines with the model (assuming a micro-turbulent line broadening of $1.0$~km\,s$^{-1}$, in addition to the thermal motion). The CO envelope size is calculated from the photodissociation model of \citet{Mamon88}. The applicability of these results to the present-day wind in these objects is a bit questionable since the shell itself may provide shielding against photodissociation. In the modelling, the CO $J=1\rightarrow0$ intensity is most sensitive
to the size of the envelope \citep{Schoeier01}. 

For some of the younger detached shell sources like \object{R~Scl}, \object{V644~Sco}, and \object{DR Ser} it is difficult, given the current set of CO observations available, to estimate reliable mass-loss rates. In these objects the observed HCN line emission has been used instead. HCN is only probing the present-day wind since it is readily photodissociated at the distances where the detached shells are found. Note, however, that there is some indication that HCN $J=1\rightarrow0$ emission from the shell is detected in \object{V644~Sco} (Fig.~\ref{spectra}). As in  the case of the CO modelling an assumption of the HCN abundance distribution needs to be made in order to estimate the mass-loss rate. HCN is a species with a photospheric origin. \citet{Olofsson93b} estimated HCN abundances in the photospheres of four of the sample stars and obtained values of $\sim 0.8 - 5.4\times10^{-5}$ relative to H$_2$. Here we adopt a generic value of $2\times10^{-5}$ for the analysis. The size of the HCN envelope was then calculated from a photodissociation model  based on the results from \citet{Lindqvist00}. It should be pointed out that for each HCN model a CO model using the same mass-loss rate needs to be calculated first in order to obtain the correct temperature profile to use. This also makes it possible to check if the CO model is consistent with the observed CO lines. In all three cases it is.

\object{U~Cam} is the only source where the abundance of HCN in the aCSE can be determined based on the results from the CO modelling. Using the mass-loss rate and expansion velocities from Table~\ref{gas}, an HCN abundance of approximately $4\times10^{-5}$ (relative to that of H$_2$) is obtained using the results from \citet{Lindqvist00} to calculate the HCN envelope size. A more detailed investigation of the HCN and CN envelopes around \object{U~Cam} based on new high-spatial-resolution interferometric observations is underway (Lindqvist et al., in prep.). U~Cam,  together with V644~Sco, are the only sources where HCN emission from the detached shell is detected.

The expansion velocities and derived mass-loss rates for the present-day wind (aCSE)  are presented in Table~\ref{gas}. The uncertainties in the mass-loss-rate estimates, within the adopted circumstellar model, are about $\pm 50\%$ for estimates based on CO observations and as high as a factor $\sim 5$ when HCN emission is used, mainly due to the larger uncertainty in the adopted photospheric HCN abundance.
Examples of line profiles taken from the best-fit models, including both contribution from the present day wind and detached shell, are given in Fig.~\ref{models} for \object{U~Cam} and \object{U~Ant}. 

\begin{figure}
\centerline{\includegraphics[width=7.5cm]{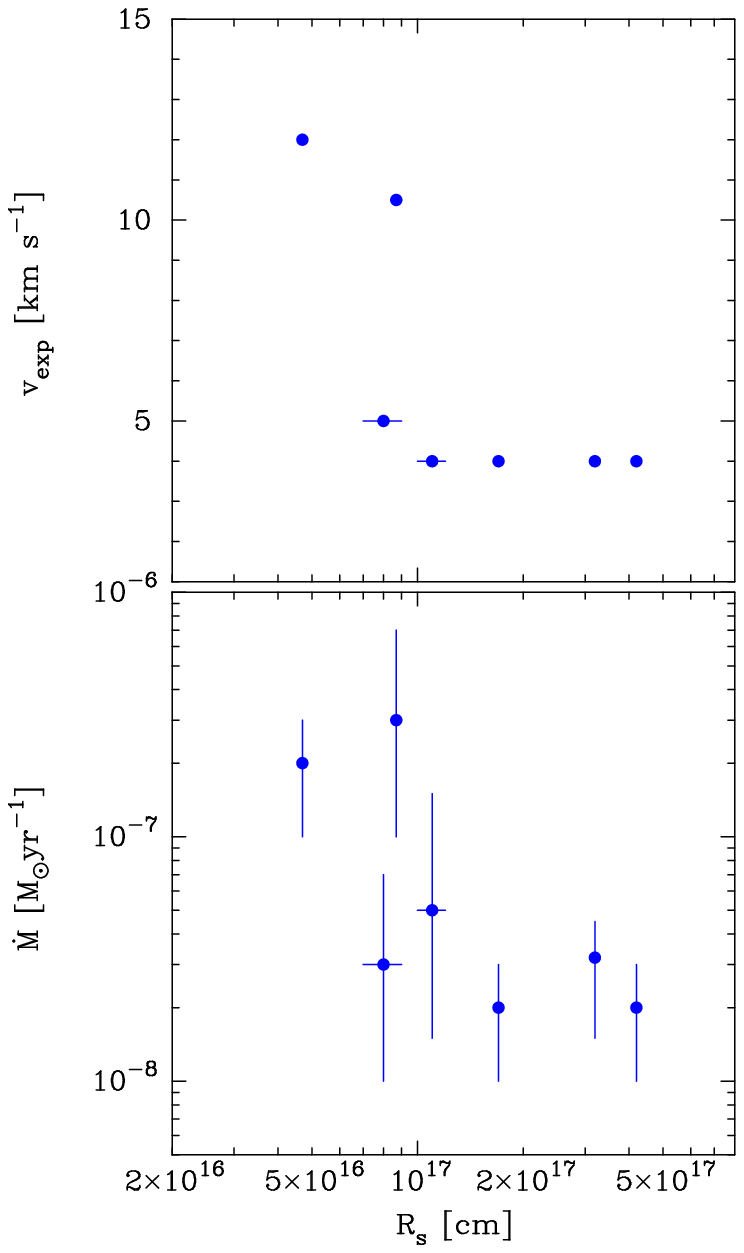}}
\caption{The present-day-wind gas expansion velocity (upper panel) and mass-loss rate (lower panel) plotted as a function of the radius of the shell ($R_{\mathrm s}$).} 
  \label{present_radius}
\end{figure}

In Fig.~\ref{present_radius} the present-day-wind mass-loss rates and gas expansion
velocities are plotted against the size of the detached shell. Even though the 
uncertainties in the mass-loss rates are substantial (particularly those based on
HCN emission) we conclude that there is a trend of decreasing present-day mass-loss
rate with increasing size of the shell.
This is further corroborated by the similar decrease in the gas expansion velocity with
increasing size of the shell. The velocity is an observational property that can
be determined with high accuracy, and it is known to correlate reasonably well
with the mass-loss rate for `normal' CSEs \citep[e.g.,][]{Olofsson03}. The
implications of this result is further discussed in Sect.~\ref{discussion}.

\subsection{The detached shell around \object{R~Scl}}
\object{R~Scl} deserves a separate discussion. From the dust modelling it appears to be a normal detached shell source (see Table~\ref{dust}). However, the CO line modelling of the main isotope places the shell at about half the distance of what is actually determined from scattered-light observations \citep{Delgado01}. The mass of the shell is also unusually high $\approx$\,1.0$\times$10$^{-2}$\,M$_{\odot}$ for  such a young shell. Modelling of the rarer $^{13}$CO isotopomer, however, gives results consistent with the larger radial position of the shell, and a much lower shell mass of approximately 2.5$\times$10$^{-3}$\,M$_{\odot}$ is obtained [adopting the photospheric $^{12}$C/$^{13}$C ratio of 19 determined by \citet{Lambert86}]. The properties of the detached shell around \object{R~Scl} reported in Table~\ref{gas} are based on the $^{13}$CO modelling.

The reason for the discrepancy obtained from the $^{12}$CO modelling is not known. Possibly, the mass-loss-rate history of this object is more complicated. Therefore, detailed interferometric CO observations are important for resolving this issue.

\begin{figure}
\centerline{\includegraphics[width=7cm]{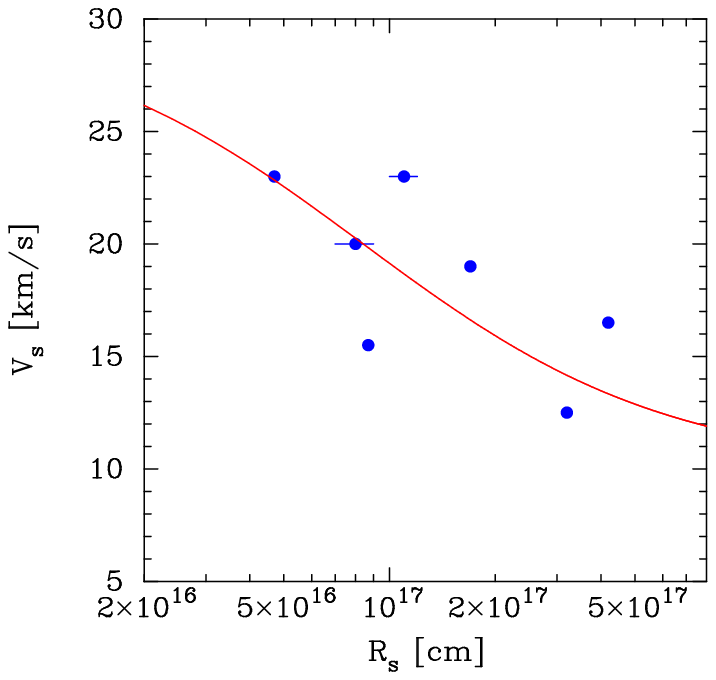}}
  \caption{The shell expansion velocity ($V_{\mathrm{s}}$) plotted against the shell radius ($R_{\mathrm s}$; as determined from CO line data). The solid line shows the expected velocity-radius relation of the interacting wind model presented in Sect.~\ref{interaction_model}}. 
\label{vel_radius}
\end{figure}
\begin{figure}
\centerline{\includegraphics[width=7cm]{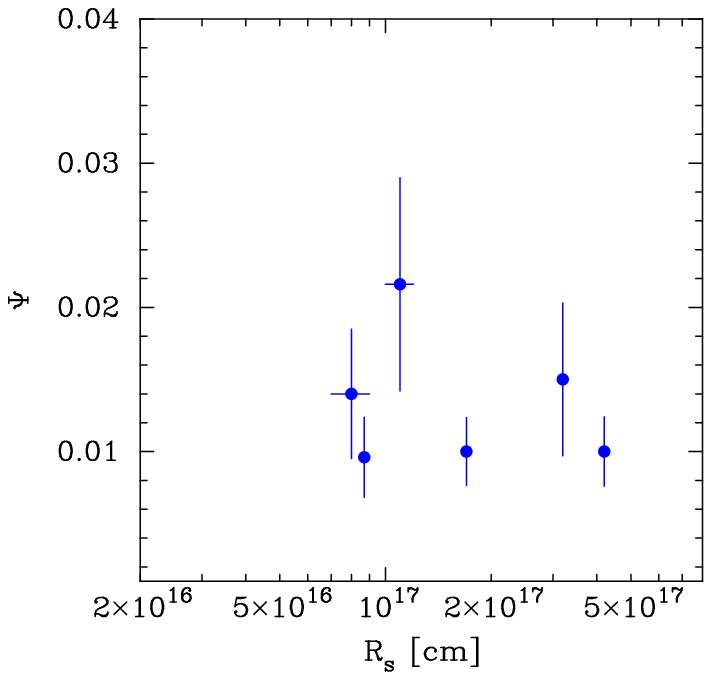}}
  \caption{The  dust-to-gas ratio ($\Psi$) plotted against the radius of the shell ($R_{\mathrm s}$; assuming that the dust and gas shells coincide spatially).} 
  \label{ratio_radius}
\end{figure}
\begin{figure}
\centerline{\includegraphics[width=7cm]{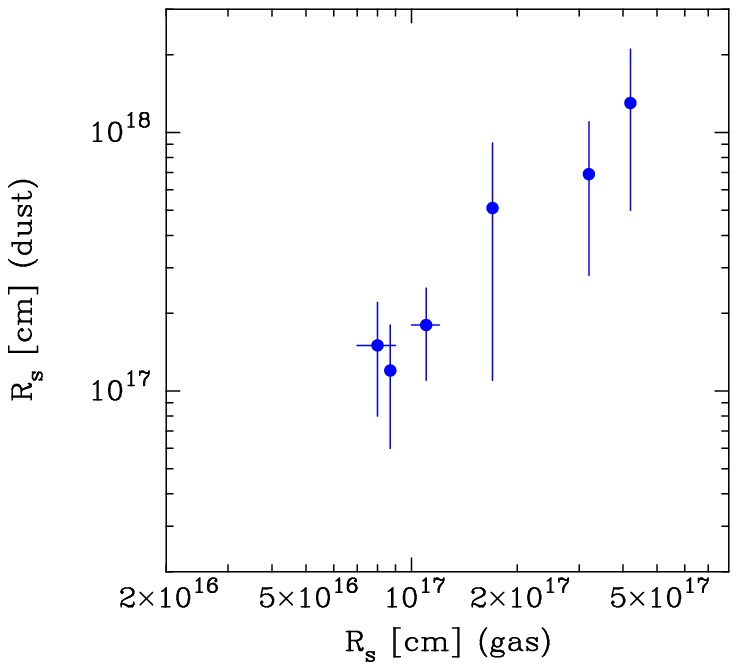}}
  \caption{Relation between the radius of the shell as determined from dust and gas modelling, respectively.} 
  \label{Rs_vs_Rs}
\end{figure}

\section{Discussion}
\label{discussion}

\subsection{Evidence for interacting winds}
\label{interaction}
As shown in Fig.~\ref{mass_radius} there is compelling evidence from both the dust and gas modelling that a detached shell gains mass as it expands away from the central star. This would indicate that material is being swept up in the process. Further support for such a scenario comes from the observed width of the CO line emission emanating from the shell. Fig.~\ref{vel_radius} shows a  correlation ($r$\,=\,$-0.64$) between the expansion velocity of the shell, $V_{\mathrm{s}}$, and its location, $R_{\mathrm{s}}$ in that the farther away the shell is located the lower its velocity. This could be naturally explained if the shell is indeed interacting with some other medium. In addition,
the high inferred kinetic temperatures (actually lower limits) are also a signpost
of interaction.

The strongest argument for the $M_{\rm s}$--$R_{\rm s}$ relation is that it is obtained in two 
completely independent ways. Nevertheless, the possibilities of systematic errors in the analysis and observational
biases should be investigated. One way of looking for possible systematic errors in the analysis, which also includes both the dust and the gas estimates, is to calculate the dust-to-gas mass-loss-rate ratio ($\Psi$) in the shells. To do this we have placed both the gas and dust shells at the same radial distance. The resulting values for $\Psi$ as a function of the radial distance of the shell are shown in Fig.~\ref{ratio_radius}. There is no apparent correlation ($r$\,$=$\,0.49) between the two quantities, and the results are consistent with the same value of $\Psi$ in all
the shells, $\approx$\,0.012. This result provides no indication of a systematic error.  On the other hand,
the dust-to-gas mass-loss-rate ratios derived for the present day winds (Table~\ref{dust}) are much lower (by a factor 3-10) than those found for the shells, but fully consistent with those  normally quoted for low to intermediate mass-loss-rate carbon stars \citep{Groenewegen98}. As shown by  \citet{Groenewegen98} there is a trend that the dust-to-gas ratio increases with mass-loss rate and that for $\dot{M}$\,$\gtrsim$\,5$\times$10$^{-6}$\,M$_{\sun}$\,yr$^{-1}$ (representative of the high-mass-loss-rate epoch  producing the detached shell) $\Psi$\,$\sim$\,0.01.

The fact that the locations of the dCSEs as determined from the dust modelling correlate very well with those estimated from CO observations, as shown in Fig.~\ref{Rs_vs_Rs}, suggests that the dust and gas shells are located relatively close spatially. \citet{Delgado03a} in their study of scattered stellar light in the CSE around \object{U~Ant} found that the dust shell is located about $7\arcsec$ further out than that of CO (corresponding to a drift velocity of $\approx$\,3\,km\,s$^{-1}$).

The most important parameter in the dust analysis is the dust temperature. The derived
dust mass scales as $T_{\rm dust}^{-1}$ in the Rayleigh-Jeans limit and as
$\exp (h\nu/kT_{\rm dust})$ in the Wien's regime, and the inferred shell radius
as $T_{\rm dust}^{-2.5}$ (if the grain emissivity scales as $\nu$). This means
that even in the case of identical shells a spread in estimated dust temperatures will result 
in a mass--radius
relation (with a scatter depending on the properties of the central stars). Thus,
in the dust analysis the accuracy of the dust temperature estimate is crucial.
As shown in Table~\ref{dust} these are rather well constrained (even
though longer-wavelength data, free of cirrus contamination, are highly
desirable), and we conclude that the dust temperature estimates are not the
reason for the mass-radius relation.

There is possibly
an observational bias in the sense that low-mass shells of large radii would be
difficult to detect in both dust and CO line emission. We cannot exclude that such shells
exist. On the other hand, we see no reason (other than the scarcity of stars with
detached shells) why high-mass shells of small size should be missed, and are
therefore reasonably certain about the upper envelope of the shell mass for a 
given shell size.

\subsection{A simple interacting-wind scenario}
\label{interaction_model}
Thus, we conclude that the observed mass-radius relations are most likely correct, 
and use a simple interacting-wind model to estimate whether our data
can be explained in such a scenario using reasonable assumptions on AGB winds. 
 \citet{Kwok78} presented the results of a fully momentum-coupled wind interaction.
In the model a fast moving wind is running into a slower moving wind. Both winds are assumed to be spherically symmetric, have constant mass-loss rates, and expand at constant velocities.
The qualitative validity of this simple picture has been verified by detailed modelling performed by  \citet{Steffen00}.
%
%
%
%
%
%

In our case it should be noted that the high velocities observed for the younger detached shell sources strongly suggests that 
the mass-loss rate forming the fast moving wind was also very high, say about 10$^{-5}$~M$_\odot$\,yr$^{-1}$, since
the mass-loss rate and gas expansion velocity correlates positively for normal
CSEs \citep[e.g.,][]{Olofsson03}. Based on Fig.~\ref{mass_radius} it is reasonable
to assume that the initial mass of the shell is of the order 10$^{-3}$~M$_\odot$,
i.e., with a mass-loss rate of 10$^{-5}$~M$_\odot$\,yr$^{-1}$ the ejection period
is about 10$^2$ years. Thus, a brief period of very high mass-loss rate is inferred producing a fast moving shell.

We have modified the model of \citet{Kwok78} to treat a scenario in which a fast moving shell is running into a slower moving wind.
Conservation of the total momentum contained in the shell (both ejected and swept-up material) gives an expression for the shell velocity $V_{\mathrm{s}}$
\begin{equation}
V_{\mathrm{s}} = \frac{M_{\mathrm{ej}} V_{\mathrm{ej}} + M_{\mathrm{sw}} V_{\mathrm{sl}}}{M_{\mathrm{ej}} + M_{\mathrm{sw}}}
\end{equation}
where $M_{\mathrm{ej}}$ is the mass contained in the initial fast moving shell at constant velocity $V_{\mathrm{ej}}$, and $V_{\mathrm{sl}}$ the constant velocity of the slower moving wind. The mass of the initial slow wind swept-up by the shell, $M_{\mathrm{sw}}$, is given by
\begin{equation}
M_{\mathrm{sw}} = \frac{\dot{M}_{\mathrm{sl}}}{V_{\mathrm{sl}}}R_{\mathrm{s}},
\end{equation}
i.e., the shell mass ($M_{\mathrm{s}} = M_{\mathrm{ej}} + M_{\mathrm{sw}}$) will increase linearly with radius. 

Assuming $V_{\mathrm{ej}}$=30\,km\,s$^{-1}$ and $M_{\mathrm{ej}}$=8$\times$10$^{-4}$\,M$_{\odot}$, further adopting the median
values for the mass-loss rate and the gas expansion velocity of a sample of
irregular and semiregular carbon stars, $\dot{M}_{\mathrm{sl}}$=3$\times$10$^{-7}$\,M$_{\odot}$\,yr$^{-1}$ and $V_{\mathrm{sl}}$=10\,km\,s$^{-1}$ \citep{Schoeier01} as typical
values for the slow wind, we find that the velocity of the shell decrease as illustrated in Fig.~\ref{vel_radius}. The mass-radius relation predicted by the model is shown in Fig.~\ref{mass_radius}. In all, this simple interacting wind model explains the observed properties of detached shells, bearing in mind that individual objects very likely have different characteristics.

Under the assumption that the slow stellar wind is that of a typical AGB star there is still the need of a very brief period of high mass loss also in  the interacting wind scenario. These eruptive events are most likely linked to a He-shell flash (thermal pulse) as originally suggested by \citet{Olofsson90}  and
\citet{Vassiliadis93}. 

It is possible that the swept-up material is depleted in CO due to it being  photodissociated to some degree. This would be most pronounced for the older shells. At the densities and temperatures prevailing in the shells there are no means of effectively producing CO \citep{Mamon88}. In this sense the estimated gas shell masses should be regarded as lower limits, in particular for the larger shells.

The cooling time scales are $\sim$\,$10^2$\,$-$\,$10^3$\,yr for gas cooling and as high as $\sim$\,$10^5$\,$-$\,$10^6$\,yr for the dust \citep{Burke83}. 
Even the gas cooling time scale is rather long compared with the shell formation and dynamical time scales. However, the time scales will be significantly reduced   in a  clumpy medium. A clumpy medium, with gas and dust well mixed, would also explain why the dust shells show effects of swept-up material despite the fact that the dust mean-free path is rather long for the densities derived assuming homogeneous shells.

Finally, we have found evidence that the present-day mass-loss rate is decreasing
with the size of the shell. The most reasonable interpretation of this is that
the mass-loss rate declines to very low levels after the brief period of very
high mass loss on a time scale of a few thousand years. Thus, there should be 
material inside the shells. There is no indication of this in the interferometer CO
line maps, suggesting that here the CO molecules have been photodissociated. 
In the
shells the densities are higher and the CO molecules survive longer. This is
supported by the detections of $^{13}$CO in some cases.
However, it should be noted that interferometers are notoriously insensitive to weak extended emission and that \citet{Lindqvist99} and \citet{Olofsson00} only recover about 50\% of the total flux in their interferometric maps.

\subsection{Photodissociation}
The $^{12}$CO/$^{13}$CO-ratio in the shells of \object{V644~Sco} and \object{S~Sct} are 25 and 20, respectively.  For \object{S~Sct} there exist photospheric estimates of the $^{12}$C/$^{13}$C-ratio.  \citet{Lambert86} have estimated this ratio to be 45, significantly larger than the value of 14 subsequently obtained by \citet{Ohnaka96} in a different analysis. The rather low values for the circumstellar ratio, and in the case of \object{S~Sct} a rough agreement with the estimated photospheric ratio, suggest that photodissociation of CO molecules is low in the shells. If photodissociation would be effective this ratio should be much higher due to the lower ability of self-shielding for the rarer $^{13}$CO isotopomer. The fact that the CO molecules appear to survive even out to distances of $\approx$\,5$\times$10$^{17}$\,cm would suggest that the medium is clumpy to a large degree, as indeed suggested by observations \citep{Olofsson96,Lindqvist99,Olofsson00}.

\subsection{Predictions for future sub-millimetre observations}
\label{sec_pred}
To date only three sources have been studied in detail using interferometry. It is of importance also to map the remaining three sources, in particular \object{R~Scl}. However, high-$J$ CO observations are also useful, in particular to constrain the kinetic gas temperature in the shell. To illustrate this, predictions for high-frequency CO observations have been calculated based on the best-fit model for \object{V644~Sco} using a 15\,m telescope. The results are presented in Table~\ref{predictions} for three different temperatures in the shell. Note that the density has been decreased slightly when the temperature was raised in order to maintain a good fit to the observed CO lines ($J=1\rightarrow0, 2\rightarrow1, 3\rightarrow2$).

These results show the potential of 
single-dish telescopes such as CSO, JCMT and the upcoming APEX{\footnote{The Atacama Pathfinder EXperiment (APEX), is a collaboration between Max Planck Institut f\"ur Radioastronomie (in collaboration with Astronomisches Institut Ruhr Universit\"at Bochum), Onsala Space Observatory and the European Southern Observatory (ESO) to construct and operate a modified ALMA prototype antenna as a single dish on the high altitude site of Llano Chajnantor.}}  to further constrain the properties of detached shells.

\begin{table}
\caption[]{Predicted CO line intensities (in K\,km\,s$^{-1}$) for the detached shell around \object{V644~Sco}.}
\label{predictions}
$$ 
\begin{array}{p{0.3\linewidth}ccccc}
\hline
\noalign{\smallskip}
\multicolumn{1}{l}{{\mathrm{Transition}}} &
\multicolumn{5}{c}{T_{\mathrm{kin}} [\mathrm{K}]}  \\
\cline{2-6}
&
\multicolumn{1}{c}{100}  &&
\multicolumn{1}{c}{200}  &&
\multicolumn{1}{c}{400}  
\\
\noalign{\smallskip}
\hline
\noalign{\smallskip}
$J=1\rightarrow 0$ &                     \phantom{0}6.5\phantom{0}     && \phantom{0}5.8\phantom{0} && \phantom{0}5.1\phantom{0} \\
$J=2\rightarrow 1$ &                      22.6\phantom{0}                       && 22.8\phantom{0} && 22.6\phantom{0}\\
$J=3\rightarrow 2$ &                      20.7\phantom{0}                     && 22.9\phantom{0} && 25.5\phantom{0} \\
$J=4\rightarrow 3$ &                     \phantom{0} 7.8\phantom{0} && 12.5\phantom{0} && 13.2\phantom{0}\\
$J=6\rightarrow 5$ &                      \phantom{0}0.58                      && \phantom{0}1.2\phantom{0} && \phantom{0}2.3\phantom{0}\\
$J=7\rightarrow 6$ &                      \phantom{0}0.15                    && \phantom{0}0.41 && \phantom{0}1.0\phantom{0}\\
$J=8\rightarrow 7$ &                     \phantom{0}0.04                     && \phantom{0}0.15 && \phantom{0}0.44\\
\noalign{\smallskip}
\hline
\end{array}
$$ \end{table}

\section{Conclusions}
We report the detection of a detached molecular shell around the carbon star \object{DR~Ser}. The properties of the shell are similar to the young shells found around the carbon stars \object{U~Cam} and \object{V644~Sco}. With \object{DR~Ser} the total number of carbon stars with molecular detached shells are seven. In fact, there is not much hope of increasing this number further since most of
the reasonably nearby AGB stars with mass loss have already been searched for
circumstellar CO radio line emission, and for the more distant objects the double-peaked
line profiles become less pronounced and detached shell sources accordingly more
difficult to identify. We therefore considered the time ripe for a systematic analysis
of the existing objects.

Based on radiative transfer modelling of both observed molecular line emission and continuum emission the properties of the dust and gas in these shells, as well as the present-day stellar mass loss have been investigated. It turns out that there is a clear trend that both the dust and gas shell masses increase with radial distance from the central stars. At the same time the velocity by which the shells recede from  the stars decreases with radial distance. The most plausible explanation for this behaviour is that the shell is sweeping up material from a surrounding medium. 

We find that an interacting wind scenario where a brief period ($\sim$\,10$^2$\,yr)
of high mass-loss rate ($\sim$\,10$^{-5}$\,M$_\odot$\,yr$^{-1}$) results in a
high-velocity shell ($\lesssim$\,30\,km\,s$^{-1}$) that expands into a previous slow,
low-mass-loss-rate AGB wind gives an adequate description of the observations.
There is also evidence that the mass-loss rate, following the short period of very
intense mass loss, decreases on a time scale of a few
thousand years to reach a low level (a few 10$^{-8}$\,M$_\odot$\,yr$^{-1}$).
The most plausible scenario for such a mass-loss-rate modulation is a He-shell flash (thermal pulse).

The gas kinetic temperature is the least well constrained parameter in the modelling. We suggest that observing high-$J$ transitions from CO using current and future single-dish telescopes should help to constrain this parameter.

Among the seven stars with known detached molecular shells, \object{R~Scl} clearly stands out. The modelling of its $^{12}$CO line emission suggests a shell located about a factor of two closer to the star compared with observations of scattered stellar light. Also the derived mass is significantly higher than obtained from dust and $^{13}$CO line modelling. This could possibly indicate that R Scl has had a significantly different mass-loss-rate history than the other detached shell sources, and this warrants further study, in particular through interferometric CO observations.

\begin{acknowledgements}
Kay Justtanont is thanked for stimulating discussions.
The authors are grateful to the Swedish research council for financial support. This research made use of data products from the Midcourse Space 
Experiment.  Processing of the data was funded by the Ballistic 
Missile Defense Organization with additional support from NASA 
Office of Space Science.  This research has also made use of the 
NASA/ IPAC Infrared Science Archive, which is operated by the 
Jet Propulsion Laboratory, California Institute of Technology, 
under contract with the National Aeronautics and Space 
Administration.
\end{acknowledgements}

\bibliographystyle{aa}

\end{document}